\documentclass[ pra, twocolumn, superscriptaddress,
nofootinbib ]{revtex4-2}

\usepackage{amsmath,amssymb,amsthm} \usepackage{amsfonts}
\usepackage{bm} \usepackage{braket} \usepackage{graphicx}
\usepackage[ colorlinks=true, citecolor=cyan, urlcolor=blue
]{hyperref} \usepackage[noabbrev]{cleveref}

\newcommand{\I}{\mathbb{I}} \newcommand{\Dphi}{\Delta\phi}
\newcommand{\nth}{n_{\mathrm{th}}}
\newcommand{\smst}{\mathrm{SMST}}
\newcommand{\etal}{\textit{et al}.\,}

\usepackage[normalem]{ulem} %For bibliography

\usepackage{xcolor,hyperref}
\definecolor{darkblue}{rgb}{0,0.0.1,0.3}
\definecolor{darkred}{rgb}{0.6,0.1,0}
\hypersetup{colorlinks,breaklinks,
        linkcolor=darkred,urlcolor=darkblue,
        anchorcolor=darkred,citecolor=
        darkred,pdfauthor=SaAr, pdftitle=SaAr.HBT}

\begin{document}

\title{Thermal activation of photon catalysis in
difference-intensity Mach-Zehnder interferometry with
squeezed thermal light}

\author{Kirtpreet Singh} \affiliation{Department of Physical
Sciences, Indian Institute of Science Education and Research
Mohali, Sector 81 SAS Nagar, Manauli PO 140306, Punjab,
India}
\author{Arvind} \affiliation{Department of Physical
Sciences, Indian Institute of Science Education and Research
Mohali, Sector 81 SAS Nagar, Manauli PO 140306, Punjab,
India}

\begin{abstract}
Realistic non-Gaussian operations, which can be implemented
by mixing a nonclassical Gaussian state with a Fock state at
a beam splitter and then heralding on a
photon-number-resolving detector, are known to improve
phase estimation in a Mach-Zehnder
interferometer. The existing studies have considered the
single-mode squeezed vacuum for difference-intensity
and parity detection, and the two-mode squeezed thermal
state for parity detection alone as the initial Gaussian
resources.  However, a real squeezer in the lab actually
produces single-mode squeezed thermal light, which has not
been studied in a coherent-plus-resource interferometer.  We
take up this case and using the phase space Wigner picture,
we obtain the heralded state produced by photon subtraction,
addition and catalysis on a single-mode squeezed thermal
resource, and evaluate the corresponding phase
sensitivities. Single-photon catalysis, which is useless on
squeezed vacuum, becomes a useful operation for any thermal
occupation above $\nth\simeq0.2$, and since it heralds with
a probability near one-half, it turns out to be the best
success-weighted operation over the whole thermal range.
In the vacuum limit our results reproduce the known
hierarchy, where two-photon catalysis is the best operation.
The quantum Fisher information computations confirms that
thermal noise enhances the phase information encoded in the
catalyzed state, demonstrating that difference-intensity
detection closely approaches the ultimate precision bound
across most operational regimes, and, together with a loss
analysis, verifies the robustness of the effect.
\end{abstract}

\maketitle

\section{Introduction} \label{sec:intro}

In the realm of optical metrology, the Mach-Zehnder
interferometer (MZI) which is equivalent to the Michelson
interferometer, is one of the most widely used devices for
measuring an unknown
phase~\cite{Caves1981,Dowling2008,Giovannetti2011}. When
only classical resources are used, the sensitivity of the
MZI cannot go below the shot-noise level
(SNL)~\cite{Caves1981}. To beat this limit and to move towards
the Heisenberg limit, various nonclassical states like
squeezed states, NOON states and Fock states have been
employed~\cite{Giovannetti2004,Boto2000,Hofmann2007,Anisimov2010,Lang2013,Kwon2019}.
This has made very precise phase measurements possible,
which are useful in many areas, including gravitational wave
detection~\cite{LIGO2019}, quantum-enhanced dark matter
searches~\cite{Backes2021}, and biological
sensing~\cite{Taylor2016}.

In particular, the single-mode squeezed vacuum (SSV) state
together with a coherent state is a very common and
practical choice of inputs to the MZI for the estimation of
an unknown
phase~\cite{Gard2017,Lang2013,Seshadreesan2011,Ataman2018,Mishra2022}.
Squeezed light was first observed by Slusher \etal in
1985~\cite{Slusher1985}, and since then it has become an
important tool in quantum metrology for improving phase
sensitivity~\cite{Vahlbruch2008,Vahlbruch2016}. Since
generating a high level of squeezing is experimentally quite
hard~\cite{Vahlbruch2016}, it is worthwhile to look for ways
to improve the phase sensitivity even with a modest amount
of squeezing. General squeezed displaced thermal states are
represented by Gaussian Wigner distributions and it is
convenient to handle them using phase space methods.
\cite{WeedbrookRMP2012,AdessoRagy2014,BraunsteinvanLoock2005,Arvind1995}.

 Non-Gaussian states generated by non-Gaussian operations
have been used in many quantum protocols such as
entanglement
distillation~\cite{Lvovsky2002,Kim2008,Takahashi2010},
quantum
teleportation~\cite{Opatrny2000,DellAnno2007,KumarArora2023},
quantum key distribution~\cite{Huang2013,Ma2018,Hu2020}, and
quantum
metrology~\cite{Braun2014,Birrittella2012,Carranza2012,Ouyang2016,Zhang2021}.
 The three most common
non-Gaussian operations are photon subtraction (PS), photon
addition (PA) and photon catalysis
(PC)~\cite{Lvovsky2002,Kim2008,LeeNha2011}. In most of the
theoretical works, PS and PA are described using the
annihilation and creation operators $\hat a$ and $\hat
a^{\dagger}$. However, these non-unitary operations
 cannot be directly implemented in a laboratory.
To get around this problem, one uses a realistic model in
which a Gaussian resource
state is mixed with a Fock state
$\ket{k}$ at a beam splitter of transmissivity $\tau$,
followed by the detection of $l$ photons at the other
output~\cite{Bartley2015,VermaKumar2023,KumarRishabh2022}.
Here $k<l$ gives subtraction, $k>l$ gives addition, and
$k=l$ gives catalysis. Such operations have already been
realized in the
laboratory~\cite{Zavatta2004,Wenger2004,Parigi2007,Ourjoumtsev2006}.
In most of these protocols one needs to work with beam
splitters of varying transmission coefficient
$\tau$~\cite{Wolf2003,Ivan2011}. Particularly for the case
of metrology with difference-intensity method the
requirement of optimization
over $\tau$ has been emphasized~\cite{Samantaray2020}.

The non-Gaussian operations implemented in this manner
are probabilistic in nature, so their
success probability has to be taken into account, because a
large sensitivity gain which succeeds very rarely is not
really useful~\cite{VermaKumar2023,KumarRishabh2022}. The
usual way is to weight the sensitivity advantage by the
heralding probability. For the squeezed vacuum resource, it
has been found that difference-intensity detection favors
two-photon operations, with two-photon catalysis being the
best once the probability is included, whereas parity
detection favors photon
addition~\cite{VermaKumar2023,KumarRishabh2022}. In that
same work single-photon catalysis was found to be useless in
both the detectors, and the quantum Cramer--Rao bound
showed that it gives no advantage  at
all~\cite{VermaKumar2023,Braunstein1994,Paris2009}. On the
other hand, for a two-mode squeezed thermal resource
for parity detection, catalysis turns out to be the optimal
operation~\cite{KumarThermal2022}.

So far the studies have used either the pure squeezed
vacuum or a two-mode squeezed thermal state. But a real
squeezer never gives out a pure squeezed vacuum. Because of
loss and extra noise, the state that actually
generated in the laboratory is a single-mode squeezed
thermal state, which has  squeezing
along with a nonzero thermal
occupation. This resource state has not been studied
in a coherent-plus-resource interferometry setup for
difference-intensity as well as for the parity
measurements.  We take up  this precise
experimentally significant case here for
difference-intensity detection scheme on MZI.

We show in this paper that for difference-intensity
detection, single-photon catalysis, which is useless on
squeezed vacuum, becomes a useful operation once the thermal
occupation crosses a certain value.  Further, it  wins in
the success-weighted comparison because  it has the success
probability near one-half.  Thus the optimal operation
crosses over from two-photon catalysis on squeezed vacuum
single-photon catalysis on squeezed thermal light. 
The single photon catalysis is the lowest-order catalysis,
needing  a single ancilla photon and a single-photon
detection. Therefore, the crossover actually moves the optimum towards
a non-Gaussian operation which is much simpler to
implement than the earlier optimal operation namely the two
photon catalysis.

In order to evaluate how the optimal operation above
compares with the theoretically best possible sensitivity
given by the standard quantum limit, we compute the quantum
Fisher
information~\cite{Braunstein1994,Paris2009,Safranek2019,Pinel2013}.
The analysis reveals three key results: first,
difference-intensity detection closely approaches the
theoretical bound for most operations; second, the thermal
switching-on of catalysis represents an intrinsic
enhancement of the phase information within the resource
rather than a detection artifact; and finally, accounting
for optical losses demonstrates that the effect remains
robust against realistic imperfections.

The paper is organized as follows. In Sec.~\ref{sec:model}
we introduce the squeezed thermal resource, the realistic
non-Gaussian operation and the interferometer, and we give
the closed-form structure of the heralded state. In
Sec.~\ref{sec:di} we evaluate the difference-intensity phase
sensitivity, fix the vacuum baseline, and report the thermal
activation of single-photon catalysis. In Sec.~\ref{sec:qfi}
we study the quantum Fisher information, the near-optimality
of difference-intensity detection, and the robustness to
loss. Finally, Sec.~\ref{sec:conc} gives the concluding
remarks. The closed-form evaluation of the heralded moments
is given in  Appendix~\ref{app:moments}.

%%%%%%%%%%%%%%%%%%%%%%%%%%%%%%%%%%%%%%%%%%%%%%%%%%%%%%
\section{Non-Gaussian operations on a squeezed thermal
resource} \label{sec:model}

\begin{figure}[t]
\centering
\includegraphics[width=0.98\linewidth]{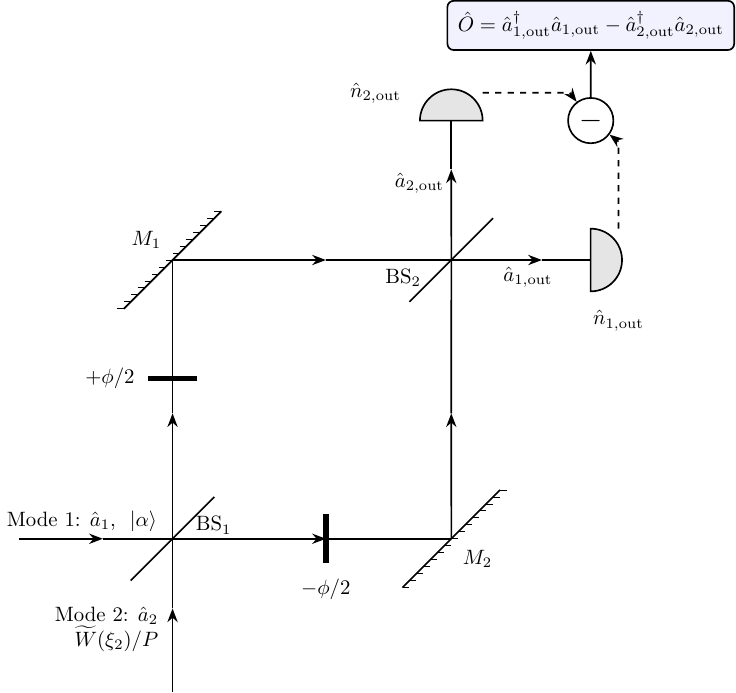}
\caption{Schematic diagram of the balanced Mach-Zehnder
interferometer (MZI) for phase estimation via
difference-intensity detection. Input mode~1 enters in a
coherent state $\ket{\alpha}$ and mode~2 is injected with
the heralded non-Gaussian squeezed thermal state described
by the normalized Wigner distribution
$\widetilde{W}(\xi_2)/P$. $\mathrm{BS}_1$ and
$\mathrm{BS}_2$ denote $50{:}50$ beam splitters, $M_1$ and
$M_2$ are highly reflective mirrors, and $\pm\phi/2$ are
phase shifters imparting a relative phase shift $\phi$
corresponding to the Schwinger rotation $e^{-i\phi
\hat{J}_2}$. The output modes $\hat{a}_{1,\mathrm{out}}$ and
$\hat{a}_{2,\mathrm{out}}$ are monitored by photodetectors
measuring photon numbers $\hat{n}_{1,\mathrm{out}}$ and
$\hat{n}_{2,\mathrm{out}}$, whose difference yields the
observable $\hat{O} = \hat{a}_{1,\mathrm{out}}^\dagger
\hat{a}_{1,\mathrm{out}} - \hat{a}_{2,\mathrm{out}}^\dagger
\hat{a}_{2,\mathrm{out}}$.} 
\label{fig:mzic} 
\end{figure}

We consider a balanced interferometer made from two
$50{:}50$ beam splitters and two phase shifters, with a coherent
state in one input port and a heralded non-Gaussian
squeezed thermal state in the other. The phase put in
by the shifters is estimated from the output.
The schematic diagram of the setup is shown in
Figure~\ref{fig:mzic} and the analysis of the interferometer
sensitivity is taken up in Section~\ref{sec:di}
In this section
we fix the resource, the operation which prepares
it, and the phase-space description of the interferometer.

\subsection{The single-mode squeezed thermal resource}
\label{sec:smst}

The single mode quantum resource states enters the
interferometer in port  $2$ which we
name mode $2$ represented by the annihilation operator
$\hat{a}_2$ with quadratures
$\xi_2=(q_2,p_2)^{T}$. A single-mode squeezed thermal state
has zero mean and the covariance matrix (with $\hbar=1$)
\begin{equation}
V_{\smst}=\kappa \begin{pmatrix} e^{-2r} & 0\\[2pt] 0 &
e^{2r}\end{pmatrix}, \qquad \kappa=\nth+\tfrac12,
\label{eq:cov}
\end{equation}
\begin{equation}
\nth = \frac{1}{\exp\!\left(\frac{\hbar \omega}{k_B T}\right) - 1},
\label{eq:planck}
\end{equation}
where $r$ is the squeezing and
$\nth$ is the mean thermal occupation.
The factor $\kappa$ represents the temperature via Planck distribution formula Eq.~\eqref{eq:planck} and for
finite temperature we have the product of the variances above the vacuum
value of
$\tfrac12$, while keeping the squeezing anisotropy
ratio the same.
Setting $\nth=0$ ($\kappa=\tfrac12$) corresponds to zero
temperature and gives back the squeezed vacuum state. The
corresponding Gaussian Wigner distribution is given by
\begin{equation} W_{\smst}(\xi_2)=\frac{1}{2\pi\kappa}\,
\exp\!\Big[-\frac{e^{2r}q_2^{2}+e^{-2r}p_2^{2}}{2\kappa}\Big],
\label{eq:Wsmst} \end{equation} which reduces to the
squeezed-vacuum Wigner function of
Refs.~\cite{VermaKumar2023,WeedbrookRMP2012} at
$\kappa=\tfrac12$. The finite thermal occupation
corresponding to a finite temperature  is the new
parameter which we keep explicit throughout the analysis.
%%%%%%%%%%%%%%%%%%%%%%%%%%%%%%%%%%%%%%%%%%%%%%%%%%%
\subsection{Realistic non-Gaussian operation}
\label{sec:ng}

\begin{figure}[t]
\centering
\includegraphics[width=0.95\linewidth]{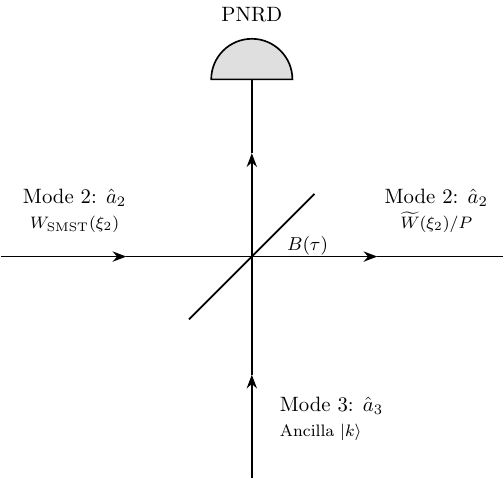}
\caption{Schematic of the realistic non-Gaussian state preparation module. An input single-mode squeezed thermal resource mode $\hat{a}_2$ described by $W_{\mathrm{SMST}}(\xi_2)$ is mixed with an ancillary mode $\hat{a}_3$ prepared in Fock state $\ket{k}$ on a beam splitter of transmissivity $\tau$, denoted by $B(\tau)$. Measuring mode~3 with a photon-number-resolving detector (PNRD) conditioned on finding $l$ photons projects the output mode~2 into the heralded non-Gaussian state $\widetilde{W}(\xi_2)/P$.}
\label{fig:mzi2}
\end{figure}

The non-Gaussian operations that we consider mix the
resource state with an ancilla prepared in a Fock state
$\ket{k}$, of the mode $\hat{a}_3$ with quadratures
$\xi_3=(q_3,p_3)^{T}$, through a beam splitter of
transmissivity $\tau$ as shown in Fig.~\ref{fig:mzi2}.
Before the beam splitter the joint Wigner function is just
the product $W_{\smst}(\xi_2)\,W_{\ket{k}}(\xi_3)$, where
the Fock-state Wigner function is:
\begin{equation}
W_{\ket{k}}(\xi_3)=\frac{(-1)^{k}}{\pi}\,
e^{-(q_3^{2}+p_3^{2})}\,L_{k}\!\big[2(q_3^{2}+p_3^{2})\big]
\label{eq:Wfock}
\end{equation}
contains the Laguerre
polynomial $L_{k}$. The beam splitter acts in the
two mode space of modes $2$ and $3$ on the quadratures,
$\xi_{23}=(\xi_2,\xi_3)^{T}$ through the orthogonal
symplectic transformation given by
\begin{equation} B(\tau)=
\begin{pmatrix}\sqrt{\tau}\,\I_2 &
\sqrt{1-\tau}\,\I_2\\[2pt] -\sqrt{1-\tau}\,\I_2 &
\sqrt{\tau}\,\I_2\end{pmatrix}, \label{eq:BS} \end{equation}
with $\I_2$ the two-dimensional identity. Heralding on $l$
photons at the ancilla output projects mode $3$ onto
$\ket{l}$ and leaves the reduced, unnormalized Wigner
function of the heralded resource as:
 \begin{equation}
\begin{split} \widetilde W(\xi_2)=2\pi\!\int d^{2}\xi_3\;
&W_{\smst}\!\big[(B^{-1}\xi_{23})_2\big]\\ &\times
W_{\ket{k}}\!\big[(B^{-1}\xi_{23})_3\big]\,
W_{\ket{l}}(\xi_3).  \end{split} \label{eq:heralded}
\end{equation}
The heralding probability is the phase-space
integral
of the unnormalized state,
\begin{equation} P=\int
d^{2}\xi_2\,\widetilde W(\xi_2), \label{eq:prob}
\end{equation} and the physical heralded resource has the
normalized Wigner function $\widetilde W(\xi_2)/P$. The
choice of $(k,l)$ selects the operation, with $k<l$ for
subtraction, $k>l$ for addition, and $k=l$ for catalysis.
Equation~\eqref{eq:heralded} has the same structure as the
squeezed-vacuum expressions of
Refs.~\cite{VermaKumar2023,KumarRishabh2022}, with the
thermal occupation coming in only through the
covariance~\eqref{eq:cov}. At $\kappa=\tfrac12$ this
construction gives back the squeezed-vacuum
results~\cite{VermaKumar2023,KumarRishabh2022}, which we use as a check in Sec.~\ref{sec:di}.

Carrying out the Gaussian integral over
the ancilla in Eq.~\eqref{eq:heralded} leaves the heralded
Wigner function in a closed form, as a finite degree
polynomial times a Gaussian,
\begin{equation}
\widetilde W(\xi_2)=\mathcal
P_{kl}(\xi_2)\,
\frac{1}{2\pi\sqrt{\det\Gamma}}\,e^{-\frac12\xi_2^{T}\Gamma^{-1}\xi_2},
\label{eq:closedW}
\end{equation}
where the diagonal envelope covariance
$\Gamma=\mathrm{diag}(\Gamma_q,\Gamma_p)$ is the
ancilla-averaged Gaussian core and the
polynomial $\mathcal P_{kl}$ of degree $2(k+l)$
is  the
non-Gaussian part. When no operation is applied the polynomial
becomes unity and the envelope gives back the bare resource
state,
with $\langle q_2^{2}\rangle_{\smst}=\kappa e^{-2r}$ and
$\langle p_2^{2}\rangle_{\smst}=\kappa e^{2r}$, and all odd
moments being identically zero.
The moments $\langle q_2^a p_2^b \rangle$ up to fourth
order, which are sufficient to evaluate both the
sensitivity and the Fisher information, are rational
functions of $e^{2r}$, $\kappa$, and $\tau$, derived in
closed form in Appendix~\ref{app:moments}.
%%%%%%%%%%%%%%%%
\subsection{Coherent and heralded resources in the
interferometer} \label{sec:input}
In this case, a  coherent state $\ket{\alpha}$ enters mode $1$ with
$\alpha= \frac{1}{\sqrt{2}}(q_0+ip_0)$ and Wigner function
\begin{equation}
W_{\ket{\alpha}}(\xi_1)=\frac{1}{\pi}\,
exp\!\big[-(q_1-q_0)^{2}-(p_1-p_0)^{2}\big].
\label{eq:coh}
\end{equation}
The input to the interferometer thus is the product
$W_{\ket{\alpha}}(\xi_1)\,\widetilde W(\xi_2)/P$. The
balanced interferometer acts through the Schwinger
generators of $\mathrm{SU}(2)$ and reduces to the rotation
$e^{-i\phi \hat J_2}$~\cite{Yurke1986,Arvind1995}, which on
phase space is given by  the symplectic transformation
\begin{equation} S(\phi)=
\begin{pmatrix}\cos(\phi/2)\,\I_2 &
-\sin(\phi/2)\,\I_2\\[2pt] \sin(\phi/2)\,\I_2 &
\cos(\phi/2)\,\I_2\end{pmatrix}, \label{eq:MZI}
\end{equation}
Therefore, the output Wigner function
is obtained by the symplectic transformation on the
quadrature operators as:
\begin{equation}
W_{\mathrm{out}}(\xi_{12})=
W_{\mathrm{in}}\!\big[S^{-1}(\phi)\,\xi_{12}\big].
\end{equation}

%%%%%%%%%%%%%%%%%%%%%%%%%%%%%%%%%%%%%%%
\section{Difference-intensity phase sensitivity}
\label{sec:di}

In this section we turn to the calculation of sensitivity of
the phase estimation via MZI described in Figure~\ref{fig:mzic} for
the resource states described in Sec.~\ref{sec:model} and make
various comparisons.
%%%%%%%%%%%%%%%%%%%%%%%%%%%%%%%%%%%%%%%%%%%%
\subsection{Sensitivity and figure of merit}
In a balanced Mach-Zehnder interferometer, the unitary
evolution of the fields under the beam splitters and the
relative phase shift $\phi$ transforms the output
photon-number difference operator $\hat{O} =
\hat{a}_{1,\mathrm{out}}^\dagger \hat{a}_{1,\mathrm{out}} -
\hat{a}_{2,\mathrm{out}}^\dagger \hat{a}_{2,\mathrm{out}}$
into the input observable:

\begin{equation}
\hat{O}(\phi) = (\hat{n}_1 - \hat{n}_2)\cos\phi + (\hat{a}_1^\dagger \hat{a}_2 + \hat{a}_2^\dagger \hat{a}_1)\sin\phi.
\label{eq:O_phi}
\end{equation}
Because our heralded resource state possesses an even Wigner distribution, all odd quadrature moments vanish ($\langle \hat{a}_2 \rangle = 0$). Consequently, the mean difference-intensity signal is directly modulated by the phase according to:
\begin{equation}
\langle \hat{O} \rangle = (\langle \hat{n}_1 \rangle - \langle \hat{n}_2 \rangle) \cos\phi.
\label{eq:mean_O}
\end{equation}
The phase sensitivity is optimal at the midpoint of the interference fringe ($\phi = \pi/2$), where the slope $|\partial \langle \hat{O} \rangle / \partial \phi| = |\langle \hat{n}_1 \rangle - \langle \hat{n}_2 \rangle|$ reaches its maximum, and the phase uncertainty follows from the standard error propagation formula:
\begin{equation}
\Dphi=\frac{\sqrt{\langle \hat O^{2}\rangle-\langle \hat
O\rangle^{2}}} {\big|\partial\langle \hat
O\rangle/\partial\phi\big|}.  \label{eq:errorprop}
\end{equation} Both the moments are symmetric-ordered
functions of the quadratures, which the Wigner picture gives
directly. For each mode,
$\hat a_i^{\dagger}\hat a_i$ and its square have the
symmetric-ordered form
\begin{eqnarray}
\hat a_i^{\dagger}\hat
a_i&=&\tfrac12(q_i^{2}+p_i^{2}-1) \nonumber \\
(\hat a_i^{\dagger}\hat a_i)^{2}&=&\tfrac14\big(q_i^{4}+p_i^{4}
-2q_i^{2}-2p_i^{2}+2\!:\!q_i^{2}p_i^{2}\!:\big).
\label{eq:nsq}
\end{eqnarray}
The output quadratures are
linear combinations of the input quadratures through
Eq.~\eqref{eq:MZI}, and the two input modes are independent,
so every average in Eqs.~\eqref{eq:errorprop}
and~\eqref{eq:nsq} splits into a coherent-mode moment, known
from Eq.~\eqref{eq:coh}, and a heralded-resource moment
coming from Eq.~\eqref{eq:heralded}. So the sensitivity only
needs the resource moments $\langle q_2^{a}p_2^{b}\rangle$
up to fourth order, together with the
probability~\eqref{eq:prob}.

Following the difference-intensity conventions of
Ref.~\cite{VermaKumar2023}, we set $q_0=10$, $p_0=0$ and
$\phi=\pi/2$, which is the phase-matching condition of the
coherent-plus-squeezed input. For each squeezing and thermal
occupation we minimize $\Dphi$ over the transmissivity
$\tau$, which is a freely tunable design parameter ~\cite{Wolf2003}. The advantage of a
heralded resource over the bare squeezed thermal resource
can be measured via a parameter $\mathcal D$ defined as follows:
\begin{equation} \mathcal
D=\Dphi^{\smst}-\Dphi^{\mathrm{NG}}, \label{eq:D}
\end{equation} and the success-weighted advantage which
decides the best operation is
obtained by multiplying $\mathcal D$  with the success probability
$P$
\begin{equation} \mathcal
R=\mathcal D \times P.
\label{eq:R}
\end{equation}

\subsection{Vacuum baseline and thermal activation of photon
catalysis}
\begin{figure}[t] \centering
\includegraphics[width=0.98\linewidth]{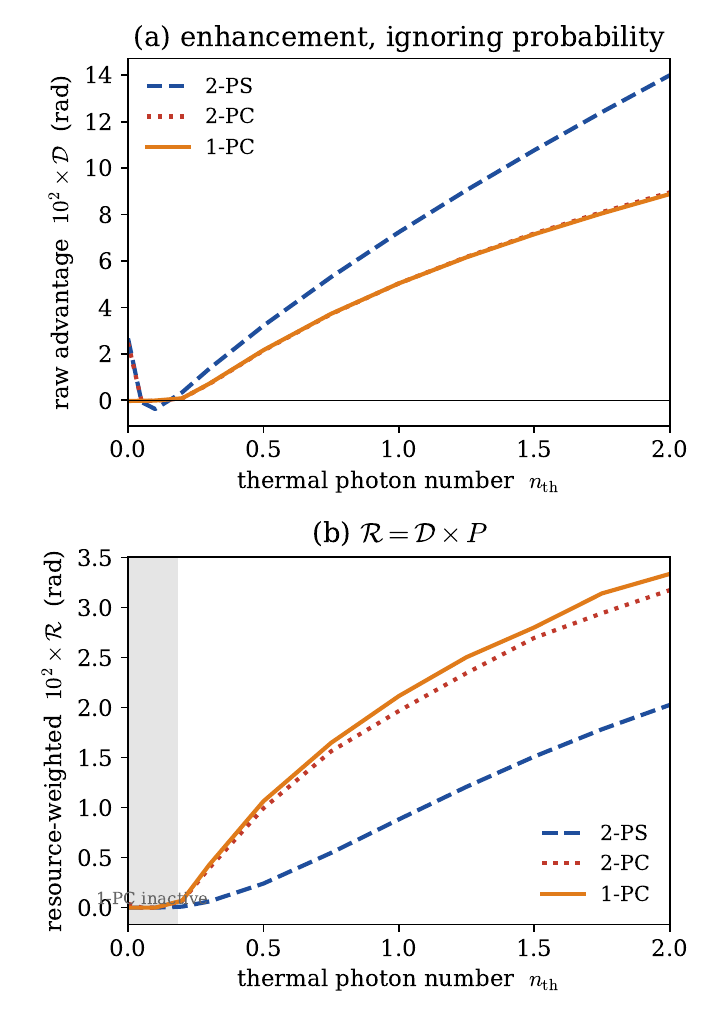}
\caption{Difference-intensity detection at $r=0.15$,
$q_0=10$, $\phi=\pi/2$, optimized over the transmissivity.
(a) Raw advantage $\mathcal D$ of Eq.~\eqref{eq:D}.
Two-photon subtraction gives the largest raw advantage over
the whole thermal range, and single-photon catalysis goes
from a useless operation at $\nth=0$ to a strongly useful
one. (b) Resource-weighted advantage $\mathcal R$
. Single-photon catalysis is the best
operation for every $\nth\gtrsim0.2$, since it works well for
heralding probability near one-half while subtraction
loses out as it has low probability of success.
}\label{fig:thermal}
\end{figure}
Our framework reproduces the known difference-intensity
results for squeezed vacuum at $\nth=0$. Single-photon
subtraction and single-photon addition do not do any better
than the bare resource, two-photon subtraction and
two-photon catalysis both improve the sensitivity at low
squeezing, and photon addition is not efficient for this
detection-scheme.
The optimal transmissivity for two-photon
catalysis comes out to be non-unit, and our value
$\tau\simeq0.13$ at $r=0.15$ agrees with the earlier
reported value.
Once the heralding probability is included  as per
Eq.~\eqref{eq:R}, two-photon catalysis is the best
difference-intensity operation on squeezed vacuum. All of
these features agree with Ref.~\cite{VermaKumar2023}.

%\subsection{Thermal activation of photon catalysis}

As we will see, thermal excitation reshuffles this
hierarchy completely.  Figure~\ref{fig:thermal} shows
the advantage $\mathcal D$ and the
success-weighted advantage $\mathcal
R$ as functions of the thermal occupation at a fixed
squeezing $r=0.15$.  Two things can be seen here. First, the
raw advantage in panel~(a) is largest for two-photon
subtraction at every thermal occupation. If one reads only
this, it would look as if subtraction is
the best operation.  But this conclusion will be misleading, because the subtraction
advantage shows up only as the transmissivity goes close to
unity, where the heralding probability drops close to
zero.
Second, single-photon catalysis behaves in a very different
way on thermal light than on vacuum. Its advantage is almost
negligible at $\nth=0$, which agrees with the fact that
single-photon catalysis is useless on squeezed vacuum,
however,
it keeps growing with the increasing thermal
occupation.
 Once the thermal occupation goes past
$\nth\simeq0.2$ it begins to outperform two
photon catalysis and the advantage reaches a value $\mathcal D\simeq0.09$~rad at
$\nth=2$. Since single-photon catalysis heralds with a
probability between $0.4$ and $0.8$, its
success-weighted
advantage in panel~(b) beats both two-photon catalysis and
two-photon subtraction for every $\nth\gtrsim0.2$. So the
best difference-intensity operation crosses over from
two-photon catalysis on vacuum to single-photon catalysis on
thermal light beyond a certain threshold temperature.

Table~\ref{tab:opt} lists the best operation at three
thermal occupations. The success-weighted advantage grows
by almost two orders of magnitude between vacuum and
$\nth=1$, which tells us that the non-Gaussian operations
help a thermal resource much more than a
zero temperature resource.
The same kind of conclusion has been
drawn for two-mode resources
under parity detection~\cite{KumarThermal2022}, and it holds
here also in the single-mode difference-intensity setting.
%%%%%%%%%%%%%%%%%%%%%%%%%%%%%%%%%%%%%%%%%%%%%%%
\begin{table}[t] \centering \caption{Best
difference-intensity operation at $r=0.15$. The advantage
$\mathcal D$, transmissivity $\tau^{\star}$, heralding
probability $P$ and success-weighted advantage $\mathcal R$
are evaluated at the transmissivity that minimizes the conditional
phase uncertainty for each operation; listed are the
one with  largest success-weighted advantage among those
operation-specific settings.} \label{tab:opt}
\begin{ruledtabular}
\begin{tabular}{ccp{1cm}ccp{1cm}}
$\nth$ & Best Operation & $\mathcal D$ (rad)& $\tau^{\star}$ & $P$ &
$10^{2}\mathcal R$ (rad)\\
%&&(rad)&&&(rad)\\
\hline $0.0$ & 2-photon cat. &
$0.025$ & $0.12$ & $0.015$ & $0.038$\\ $0.5$ & 1-photon
cat. & $0.022$ & $0.69$ & $0.492$ & $1.063$\\ $1.0$ &
1-photon cat. & $0.050$ & $0.72$ & $0.420$ & $2.115$\\
\end{tabular} \end{ruledtabular} \end{table}

\subsection{Squeezing dependence}

\begin{figure}[t] \centering
\includegraphics[width=0.98\linewidth]{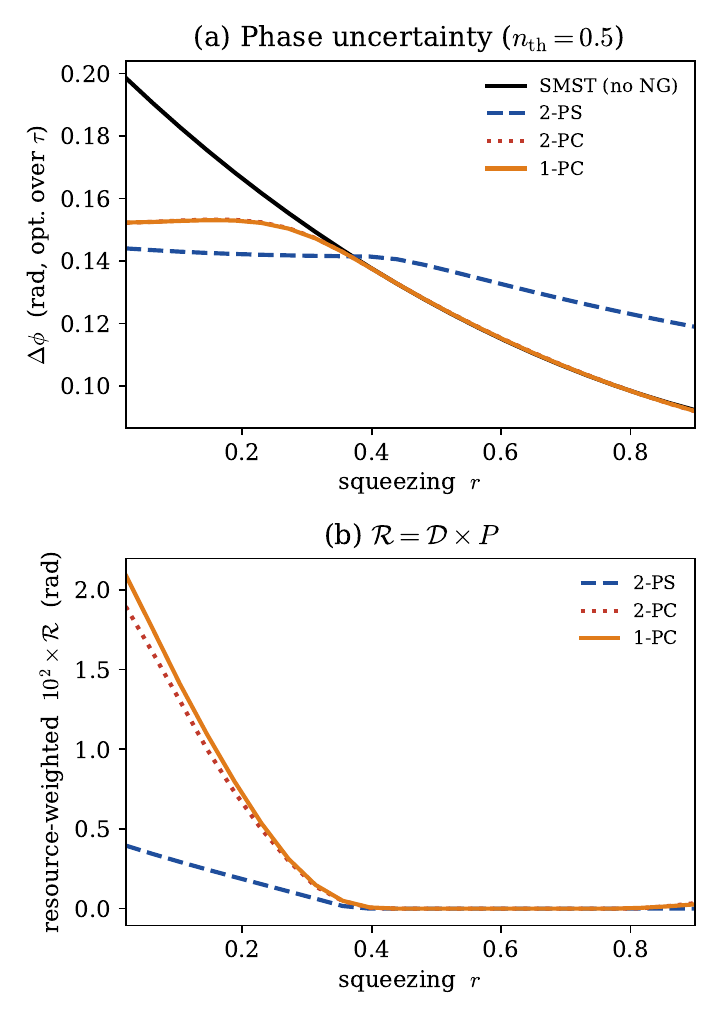}
\caption{Optimized difference-intensity phase uncertainty
$\Dphi$ against the squeezing $r$ at $\nth=0.5$, $q_0=10$,
$\phi=\pi/2$. The bare squeezed thermal resource is the
solid black curve. Single-photon catalysis improves the
sensitivity over the whole displayed squeezing window.}

\label{fig:dphir} \end{figure}

To observe the dependence on squeezing, we study the
difference-intensity measurement performance for a fixed thermal
occupation and vary the squeezing parameter for different
non-Gaussian operations on the thermal resource state.
In
Figure~\ref{fig:dphir}  the thermal occupation
is fixed at
$\nth=0.5$ and plots the optimized
phase uncertainty as a function of
the squeezing.
The performance with
the bare squeezed thermal resource keeps improving with
squeezing, as one expects.
Single-photon catalysis stays below the bare resource
case over the entire range of of squeezing parameter,
demonstrating that the activated
operation is useful over a broad squeezing window and it substantially
outperforms all the considered resources for low squeezing.
Two-photon catalysis is
competitive at the smallest squeezing and two-photon
subtraction reaches the lowest raw uncertainty at larger
squeezing, in line with the raw-advantage ordering of
Fig.~\ref{fig:thermal}(a), but neither of them carries the
heralding probability that makes single-photon catalysis the
success-weighted best.

%\subsection{Physical picture}
One can intuitively  understand why the non-Gaussian
operation of single-photon
catalysis works better for thermal resource states compared
to states at zero temperature.
Catalysis process heralds on the same photon number as
is mixed at the beam splitter with the input state,
therefore, for a pure squeezed vacuum at high transmissivity it acts
almost like doing nothing, and it hardly changes the state
and the
photon-number distribution.
A thermal resource, on the other hand, has a broader spread of
photon numbers set by $\nth$, and since the catalysis herald
picks out a particular photon number, it has  more room to operator and reshape
photon number distribution.
This reshaping makes the effective photon-number spread
that feeds the difference-intensity signal narrower and
brings down the measured variance.
Further, since the thermal resource state has a good
number of photons the success probability for heralding is
quite significant.
Subtraction and addition, on the other hand,
essentially shift the
mean photon number, and they give their advantage only when
the transmissivity goes close to one, where the herald
probability
becomes very rare.small.

The thermal occupation effectively provides the state
with the required phase-space structure for effective catalysis,
notably circumventing the vanishing-probability overhead
inherent to the other two protocols. This
enhancement originates from the mixed nature of the initial
resource state and has no counterpart in the pure vacuum
limit.

\section{Quantum Fisher information} \label{sec:qfi}

Difference-intensity detection is just one of the many
possible measurements. The quantum Fisher information gives
the best phase precision that any measurement can reach, so
it is useful to check two things. One is how close the
difference-intensity sensitivity is to this best possible
value for the same probe. The other is how much phase
information the engineered non-Gaussian resource itself
carries, and how this changes with thermal excitation.

\subsection{Interferometric bound and detector optimality}

For a pure input $\ket{\mathrm{in}}$ to the interferometer
the phase information is set by the fluctuations of the
Schwinger generator $\hat J_2$ which imprints the phase,
$F_Q=4\big(\langle\hat J_2^{2}\rangle-\langle\hat
J_2\rangle^{2}\big)$, and the quantum Cramer--Rao bound
reads
$\Dphi\ge1/\sqrt{F_Q}$~\cite{Braunstein1994,VermaKumar2023,Paris2009}.
On our coherent-plus-resource input this reduces to a moment
expression, \begin{equation} F_Q=-\tfrac12+\langle
q_1^{2}p_2^{2}\rangle+\langle p_1^{2}q_2^{2}\rangle
-2\langle{:}q_1p_1{:}\rangle\langle{:}q_2p_2{:}\rangle,
\label{eq:qfiint} \end{equation} which splits into the
coherent-mode and heralded-resource moments we have already
put together, since the two input modes are independent. At
$\nth=0$ the input is pure and Eq.~\eqref{eq:qfiint} is
exact.

Table~\ref{tab:qfi} shows the difference-intensity 
sensitivity against this bound on squeezed vacuum and
squeezed vacuum after various non-Gaussian operations.
Difference-intensity detection almost reaches the bound
for the bare resource and for two-photon subtraction,
two-photon catalysis and single-photon catalysis, so for
these operations the simple and cheap detector is close
to the best possible. Single-photon subtraction is the one
exception. Its Fisher information is the largest in the ta-
ble, but difference-intensity detection gets only a fraction
of it, which means the phase information sits in a form
that the photon-number difference does not read out well.
So difference-intensity detection is a close-to-best choice
for most of the operations, and single-photon subtrac-
tion is the one case where one would need some other
detection scheme to fully harness the available phase in-
formation.

\begin{table}[t] \centering \caption{Difference-intensity
sensitivity against the interferometric quantum
Cramer--Rao bound $1/\sqrt{F_Q}$  on squeezed vacuum at $r=0.3$,
$q_0=10$, at the representative transmissivity of each
operation. The bound is exact at $\nth=0$.} \label{tab:qfi}
\begin{ruledtabular}
 \begin{tabular}{lccc}
Operation &
$\tau$ & $1/\sqrt{F_Q}$(rad) & $\Dphi^{\mathrm{DI}}$(rad)\\
\hline Squeezed vacuum & --- & $0.105$ &
$0.105$\\ 1-photon subtraction & $0.60$ & $0.068$ &
$0.211$\\ 2-photon subtraction & $0.60$ & $0.086$ &
$0.091$\\ 2-photon catalysis & $0.15$ & $0.085$ &
$0.091$\\ 1-photon catalysis & $0.70$ & $0.137$ &
$0.138$\\ \end{tabular} \end{ruledtabular} \end{table}
%%%%%%%%%%%%%%%%%%%%%%
\subsection{Resource information and thermal activation}
%%%%%%%%%%%%%%%

To pull out the
phase information carried by the engineered resource by
itself, we look at the single-mode Fisher information of the
heralded state for a phase added
by its own number operator
\begin{equation}
\hat\rho(\phi)=e^{-i\phi\hat
n}\hat\rho\,e^{i\phi\hat n}.
\end{equation}
Since we have
$\Dphi\ge1/\sqrt{F_\phi}$ we need to calculate
$F_\phi$. For the bare squeezed thermal
resource which is Gaussian,  its Fisher information can be
written down in a closed form from the Gaussian-state expressions
of Ref.~\cite{Safranek2019}. The heralded resource is
non-Gaussian, so we get compute its Fisher information from the
spectral form ~\cite{Braunstein1994,Paris2009} as:
\begin{equation}
F_\phi=2\sum_{m,n}\frac{(\lambda_m-\lambda_n)^{2}}{\lambda_m+\lambda_n}\,
\big|\langle m|\hat n|n\rangle\big|^{2}, \label{eq:sld}
\end{equation} where
$\hat\rho=\sum_m\lambda_m\ket{m}\bra{m}$ is the number-basis
spectral decomposition of the heralded state.

When the
resource is squeezed vacuum the heralded state is pure and
Eq.~\eqref{eq:sld} becomes  the photon-number variance,
\begin{equation} F_\phi=4\big(\langle\hat
n^{2}\rangle-\langle\hat n\rangle^{2}\big),
\label{eq:qfipure} \end{equation} which is a  closed-form
expression in terms of the
fourth-order moments of Eq.~\eqref{eq:closedW}. The thermal
resource has no such  form involving
only the moments, and
Eq.~\eqref{eq:sld} is evaluated in the number basis.

\begin{figure}[t]
\centering
\includegraphics[width=0.98\linewidth]{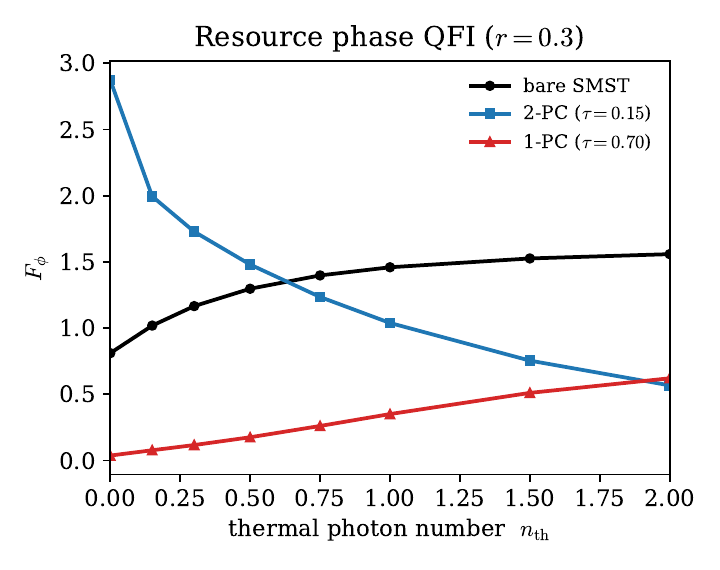}
\caption{Phase quantum Fisher information $F_\phi$ of the heralded single-mode resource as a function of thermal occupation $\nth$ at $r=0.3$, evaluated using Eq.~\eqref{eq:sld}. While two-photon catalysis and the bare squeezed thermal resource carry more phase information but decrease with thermal noise, single-photon catalysis is ineffective on vacuum and actively switches on with thermal occupation.}
\label{fig:qfi_nth}
\end{figure}

Figure~\ref{fig:qfi_nth} shows plots of the
Fisher information as a function of the thermal occupancy for
the bare squeezed
vacuum, two photon catalyzed and one photon catalyzed
resource states.

Single-photon catalysis advantage is
switched on by the thermal noise at the level of the Fisher
information itself. As is seen in the graph for this case
the value of  the Fisher information  going up from
$F_\phi\simeq0.04$ on vacuum, where catalysis is useless, to
$F_\phi\simeq0.35$ at $\nth=1$. So the advantage
 seen for
difference-intensity detection is not just a feature of that
detector, but is a real increase in the phase information
carried by the catalyzed resource.  Two-photon catalysis
instead has the largest Fisher information on vacuum and
goes down with thermal occupation. While two-photon catalysis and the bare SMST maintain a higher intrinsic Fisher information than single-photon
catalysis over the displayed range, single-photon catalysis is unique in that
its phase information is actively enhanced by thermal noise.

The Fisher information measures the phase information
contained in the state, while the
success-weighted difference-intensity advantage of
Sec.~\ref{sec:di} measures how well a
particular detection scheme
turns that information into a phase estimate. Single-photon
catalysis is the best difference-intensity operation on
thermal light, but it does not give the most Fisher
information, because catalysis reshapes the resource in a
way that suits the difference-intensity observable, and not
in the way that gives the most total phase information. So
the two ``best'' operations are different, and the
difference-intensity advantage is as much a statement about
the detector as it is about the state.

\begin{figure}[t]
\centering
\includegraphics[width=0.98\linewidth]{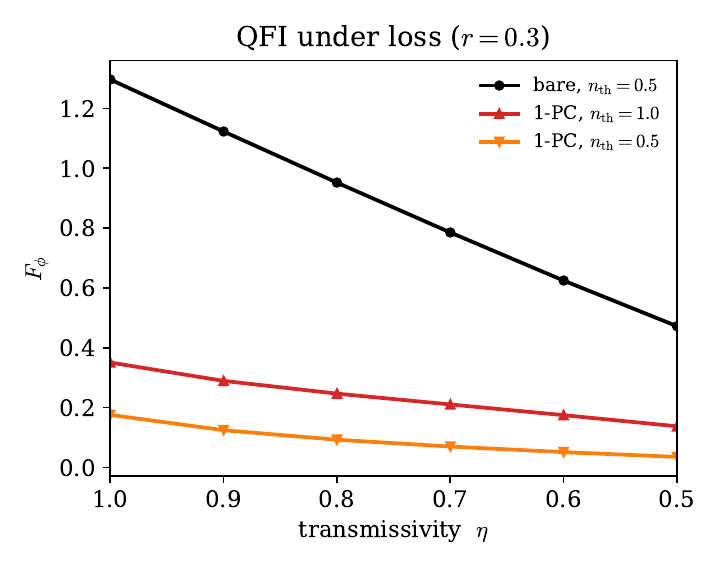}
\caption{Phase quantum Fisher information $F_\phi$ of the heralded resource under a pure-loss channel of transmissivity $\eta$. The thermal activation of single-photon catalysis survives moderate optical loss, demonstrating the robustness of the effect.}
\label{fig:qfi_loss}
\end{figure}
\subsection{Robustness against loss}

To study the robustness against loss, we introduce a pure-loss channel of transmissivity $\eta$ which acts on the heralded resource before detection. The channel is modeled by coupling the signal mode $\hat{a}$ to an environmental vacuum mode $\hat{a}_{\mathrm{env}}$ via a beam splitter of transmission efficiency $\eta \in [0, 1]$, such that $\hat{a} \to \sqrt{\eta}\,\hat{a} + \sqrt{1-\eta}\,\hat{a}_{\mathrm{env}}$. In terms of the density matrix, this corresponds to the standard photon-loss map:
\begin{equation}
\hat{\rho}_\eta = \sum_{k=0}^{\infty} \frac{(1-\eta)^k}{k!} \eta^{\hat{n}/2} \hat{a}^k \hat{\rho} (\hat{a}^\dagger)^k \eta^{\hat{n}/2} \label{eq:density}
\end{equation}

In Fig.~\ref{fig:qfi_loss} we plot the Fisher information as a function of the loss parameter $\eta$. Every resource loses phase information as the loss increases corresponding to $\eta$ going down. At $\eta = 0.75$, the catalyzed resource keeps a Fisher information at $\nth = 1$ which is close to three times its value at $\nth = 0.5$. This clearly establishes that the thermal activation by adding thermal photons to the source is useful and noise robust.

\section{Conclusion} \label{sec:conc}
In this paper we have studied the
difference-intensity phase
estimation with a coherent beam and a heralded single-mode
squeezed thermal resource state using the MZI setup.
This is the state that a real squeezer
actually  generates in the lab, and
it  had not been treated in a
coherent-plus-resource interferometer before. Working in the
Wigner picture, we obtained the heralded state after the non-Gaussian operations of photon
subtraction, addition and catalysis. The closed form
expression turns out to be a product of a finite degree
polynomial and a Gaussian.
This
allowed us to  evaluate the phase sensitivity
along with the heralding probability analytically.

Thermal excitation reshuffles the hierarchy, the
single-photon catalysis, which is useless on squeezed vacuum
and which the vacuum Cramer--Rao bound rules out under
every measurement, becomes a useful operation for any
thermal occupation above $\nth\simeq0.2$. Further, since it
heralds with a success probability near one-half it is the best
success-weighted operation over the whole thermal range.

The best operation thus crosses over from two-photon catalysis on
vacuum to single-photon catalysis on thermal light and interestingly, is a movement towards a lower
order operation which is easier to implement.
The cross-over happens, because the number-selective
catalysis heralding has a broad thermal photon-number
distribution to reshape, while the mean-shifting operations
get their advantage only where the herald becomes
very less probable.

The quantum Fisher information calculations puts the results
on a firmer footing and confirms that the thermal noise
raises the phase information carried by the catalyzed
resource state.
It turns out that
difference-intensity detection is close to the best possible
phase detection scheme for all the cases other than
single-photon subtraction for which one needs to think of a
better detection scheme. The thermal activation also
increases the noise robustness of the detection scheme.

There are two natural directions to take this work
further. Parity detection would complete the analysis as it is
the most natural other detection that one can imagine in
this direction. The vacuum version, with the
phase-matching condition $p_0=0$ and $\phi\to0$ at a
moderate coherent amplitude, is already known and works
through the same Wigner machinery~\cite{VermaKumar2023}.
The moment-generating structure of the heralded state also
supports homodyne and quadratic-homodyne
detection~\cite{KumarRishabh2022}. Finally, it would
also be interesting to see how these results change when the
resource evolves in a dissipative
environment~\cite{Rishabh2022,Olivares2012,Gurvir2025}, and to extend
the analysis to genuinely multimode resources.

\section*{Acknowledgements}
K.S. would like to specifically thank Gurvir Singh for insightful discussions, valuable ideas, and constructive feedback throughout the course of this work.
%%%%%%%%%%%%%%%%%%%%%%%%%%%%%%%%%%%%%%%%%%%
\appendix
\section{Closed-form evaluation of the heralded moments}
\label{app:moments}

We provide here the full
derivation of the unnormalized heralded Wigner function
given in Eq.~(7), explicit expressions for
the envelope covariance, and the quadrature
scheme used to evaluate the heralded moments up to fourth
order.

\subsection{Phase-space integration and covariance
completing-the-square}

The single-mode squeezed thermal resource state enters mode~2 with
covariance matrix $V_{\mathrm{SMST}} = \kappa
\operatorname{diag}(e^{-2r}, e^{2r})$, where $\kappa =
n_{\mathrm{th}} + 1/2$. Mode~3 contains an ancillary Fock
state $|k\rangle$, whose Wigner function is \begin{equation}
W_{|k\rangle}(\xi_3) = \frac{(-1)^k}{\pi} e^{-\|\xi_3\|^2}
L_k(2\|\xi_3\|^2), \label{eq:app_wigner_fock} \end{equation}
with $\|\xi_3\|^2 = q_3^2 + p_3^2$, and $L_k(\cdot)$
denoting the $k$-th Laguerre polynomial.

The beam splitter of transmissivity $\tau$ action is through the symplectic transformation
 on the quadratures
$\boldsymbol{\xi}_{23} = (\xi_2, \xi_3)^T$: \begin{equation}
\begin{pmatrix} \xi_2' \\ \xi_3' \end{pmatrix} = B(\tau)
\begin{pmatrix} \xi_2 \\ \xi_3 \end{pmatrix} =
\begin{pmatrix} \sqrt{\tau}\,\mathbb{I}_2 &
\sqrt{1-\tau}\,\mathbb{I}_2 \\ -\sqrt{1-\tau}\,\mathbb{I}_2
& \sqrt{\tau}\,\mathbb{I}_2 \end{pmatrix} \begin{pmatrix}
\xi_2 \\ \xi_3 \end{pmatrix}.
\end{equation}
Conditioned on projecting mode~3 onto the output Fock state $|l\rangle$,
the unnormalized Wigner function of the remaining state in
mode~2 is given by
\begin{equation}
\begin{split}
\tilde{W}(\xi_2) = 2\pi \int_{\mathbb{R}^2} d^2\xi_3 \,
&W_{\mathrm{SMST}}\left(\sqrt{\tau}\xi_2 - \sqrt{1-\tau}\xi_3\right) \\
&\times W_{|k\rangle}\left(\sqrt{1-\tau}\xi_2 + \sqrt{\tau}\xi_3\right) W_{|l\rangle}(\xi_3).
\end{split}
\label{eq:app_unnorm_wigner}
\end{equation}
Upon substituting Eqs.~\eqref{eq:Wsmst} and~\eqref{eq:Wfock}, the exponential terms in the integrand gather into $\exp[-\mathcal{Q}(\xi_2, \xi_3)]$, where $\mathcal{Q}$ is the quadratic form:
\begin{equation}
\begin{split}
\mathcal{Q}(\xi_2, \xi_3) &= \frac{1}{2} (\sqrt{\tau}\xi_2 -
\sqrt{1-\tau}\xi_3)^T V_{\mathrm{SMST}}^{-1}
(\sqrt{\tau}\xi_2 - \sqrt{1-\tau}\xi_3) \\
&\quad + \|\sqrt{1-\tau}\xi_2 + \sqrt{\tau}\xi_3\|^2 + \|\xi_3\|^2.
\end{split}
\end{equation}
Because $V_{\mathrm{SMST}}$ is
diagonal, the $q$- and $p$-quadratures decouple completely
into identical independent structures differing only by the
substitution $e^{-2r} \leftrightarrow e^{2r}$. We can thus
express $\mathcal{Q}$ in block form:
\begin{equation}
\mathcal{Q}(\xi_2, \xi_3) = \frac{1}{2}\xi_2^T M_{22} \xi_2
+ \xi_3^T M_{32} \xi_2 + \frac{1}{2}\xi_3^T M_{33} \xi_3,
\end{equation}
with the diagonal blocks defined by
\begin{align} M_{22} &= \tau V_{\mathrm{SMST}}^{-1} +
2(1-\tau)\mathbb{I}_2, \nonumber
\\ M_{33} &=
(1-\tau)V_{\mathrm{SMST}}^{-1} + (1+\tau)\mathbb{I}_2,
\nonumber \\
M_{32} &= M_{23}^T = -\sqrt{\tau(1-\tau)}\left(
V_{\mathrm{SMST}}^{-1} - \mathbb{I}_2 \right).  \end{align}

Completing the square over the ancilla coordinate $\xi_3$ by
shifting $\boldsymbol{u} = \xi_3 + M_{33}^{-1} M_{32} \xi_2$
gives the effective envelope covariance $\Gamma =
\operatorname{diag}(\Gamma_q, \Gamma_p)$ for the heralded
mode~2:
\begin{equation}
\Gamma^{-1} = M_{22} - M_{32}^T
M_{33}^{-1} M_{32}.  \label{eq:app_gamma_def}
\end{equation}
Carrying out the matrix inverses along each
quadrature axis gives the closed-form envelope variances:
\begin{align} \Gamma_q &= \frac{\kappa e^{-2r}(1-\tau) +
\tau}{\tau \kappa e^{-2r} + (1-\tau)},
\label{eq:app_gamma_q} \\ \Gamma_p &= \frac{\kappa
e^{2r}(1-\tau) + \tau}{\tau \kappa e^{2r} + (1-\tau)}.
\label{eq:app_gamma_p} \end{align} At zero thermal
occupation ($\kappa = 1/2$), Eqs.~\eqref{eq:app_gamma_q} and
\eqref{eq:app_gamma_p} reduce exactly to the squeezed-vacuum
envelope derived in Ref.~[40].

\subsection{Polynomial prefactor structure}

Because the Laguerre polynomials expand into finite series,
\begin{equation} L_n(x) = \sum_{m=0}^n \frac{(-1)^m}{m!}
\binom{n}{m} x^m, \end{equation} the product
$L_k(2\|\sqrt{1-\tau}\xi_2 + \sqrt{\tau}\xi_3\|^2)
L_l(2\|\xi_3\|^2)$ is a polynomial in $(\xi_2, \xi_3)$ of
total degree $2(k+l)$. Integrating this polynomial against the Gaussian kernel $\exp(-\frac{1}{2}\boldsymbol{u}^T M_{33} \boldsymbol{u})$ over the shifted coordinate $\boldsymbol{u}$ defined above gives an unnormalized heralded Wigner function of the factored form:
\begin{equation}
\tilde{W}(\xi_2) =
\mathcal{P}_{kl}(\xi_2) \frac{1}{2\pi \sqrt{\det \Gamma}}
\exp\left( -\frac{1}{2}\xi_2^T \Gamma^{-1} \xi_2 \right),
\label{eq:app_factored_wigner}
\end{equation}
where
$\mathcal{P}_{kl}(\xi_2)$ is an even polynomial in $(q_2,
p_2)$ of degree $2(k+l)$.

For single-photon catalysis ($k=1, l=1$), the
corresponding
polynomial takes the form: \begin{equation}
\mathcal{P}_{11}(q_2, p_2) = c_{00} + c_{20} q_2^2 + c_{02}
p_2^2 + c_{40} q_2^4 + c_{04} p_2^4 + c_{22} q_2^2 p_2^2,
\end{equation} where the coefficients $c_{ij}$ are rational
functions of $\kappa$, $\tau$, and $e^{\pm 2r}$.

\subsection{Exact evaluation via Gauss-Hermite quadrature}

The heralded normalization (probability of success) $P$ and
the symmetrical quadrature moments $\langle q_2^a p_2^b
\rangle$ are given by \begin{eqnarray} P &=&
\int_{\mathbb{R}^2} d^2\xi_2 \, \tilde{W}(\xi_2), \nonumber
\\
\langle q_2^a p_2^b \rangle &=& \frac{1}{P}
\int_{\mathbb{R}^2} d^2\xi_2 \, q_2^a p_2^b \,
\tilde{W}(\xi_2).
\end{eqnarray}
By symmetry, odd moments
vanish identically. To evaluate the even moments efficiently
up to fourth order ($a + b \le 4$), we transform the
integrals to the dimensionless coordinates:
\begin{equation}
u = \frac{q_2}{\sqrt{2\Gamma_q}}, \qquad v
= \frac{p_2}{\sqrt{2\Gamma_p}}.
\end{equation}
With this
substitution, the integral becomes
\begin{equation}
\begin{split} \langle q_2^a p_2^b \rangle &=
\frac{(2\Gamma_q)^{a/2} (2\Gamma_p)^{b/2}}{\pi P}
\iint_{\mathbb{R}^2} du\,dv \, e^{-(u^2 + v^2)} u^a v^b \\
&\quad \times \mathcal{P}_{kl}\left(\sqrt{2\Gamma_q} u,
\sqrt{2\Gamma_p} v\right).  \end{split}
\label{eq:app_gh_integral}
\end{equation}
Equation~\eqref{eq:app_gh_integral} matches the canonical weight function $w(x) = e^{-x^2}$ of Gauss-Hermite quadrature. For an $N_g$-point one-dimensional Gauss-Hermite quadrature formula with nodes $x_i$ and weights $w_i$, any polynomial of degree up to $2N_g - 1$ is integrated exactly without numerical truncation error:
\begin{equation}
\int_{-\infty}^\infty dx \, e^{-x^2} f(x) = \sum_{i=1}^{N_g} w_i f(x_i), \quad \text{for } \deg(f) \le 2N_g - 1.
\end{equation}
For an operation with parameter $(k,l)$, the
integrand in Eq.~\eqref{eq:app_gh_integral} along either
coordinate has maximum polynomial degree: \begin{equation}
D_{\max} = 4 + 2(k+l).  \end{equation} Thus, setting the
number of quadrature nodes such that \begin{equation} N_g
\ge \frac{D_{\max} + 1}{2} = k + l + \frac{5}{2}
\end{equation} guarantees that the numerical integration is
mathematically exact to machine precision. For single-photon
catalysis ($k=l=1$, $D_{\max}=8$), choosing $N_g = 5$ nodes
along each quadrature dimension is sufficient for exact
convergence; for two-photon catalysis ($k=l=2$,
$D_{\max}=12$), $N_g \ge 7$ ensures zero numerical
quadrature error. In our computational pipeline, we
conservatively choose $N_g = 10$, evaluating all moments via
the double sum: \begin{equation} \begin{split} \langle q_2^a
p_2^b \rangle &=
\frac{(2\Gamma_q)^{a/2}(2\Gamma_p)^{b/2}}{\pi P}
\sum_{i=1}^{N_g} \sum_{j=1}^{N_g} w_i w_j \, u_i^a v_j^b \\
&\quad \times \mathcal{P}_{kl}\left(\sqrt{2\Gamma_q} u_i,
\sqrt{2\Gamma_p} v_j\right).  \end{split}
\end{equation}
These moments are what we need to calculate
the sensitivity
$\Delta\phi$ and the quantum Fisher information.
\bibliography{references}

\end{document}